\documentclass[a4paper,12pt]{article}
\usepackage{jheppub} 

\title{\boldmath Quasinormal bulk-edge characters  of gravitons in Nariai geometry }

\usepackage{comment}

\author{Jyotirmoy Mukherjee}
\affiliation[a]{ Department of Theoretical Physics\\
Tata Institute for Fundamental Research, Mumbai 400005, India.}
\emailAdd{jyotirmoy.mukherjee\_119@tifr.res.in}

\abstract{In this paper, we evaluate the graviton character partition function in the Nariai geometry using the quasinormal mode spectrum. The character partition function obtained from the quasinormal modes via the Denef--Hartnoll--Sachdev (DHS) prescription defines the bulk contribution to the one-loop determinant. The edge partition function is then computed by subtracting this bulk contribution from the full one-loop partition function on the Euclidean continuation of the Nariai geometry, namely \(S^2\times S^2\). We find that the resulting edge partition function can be interpreted as a path integral over lower-spin fields localized on the codimension-two surface.
 }

\begin{document}
\maketitle
\flushbottom
\section{Introduction}

 Quantum corrections to black hole spacetimes offer crucial insights into the microscopic origin of black hole entropy. In particular, one-loop corrections to the Euclidean partition function of a black hole give rise to logarithmic corrections to the Bekenstein–Hawking entropy, providing a non-trivial precision test of the microscopic origin of black hole entropy \cite{Banerjee:2010qc,Banerjee:2011jp}. At the semi-classical level, such corrections are encoded in the one-loop partition function of  fields including the matter fields,  gauge fields, gravitons and the corresponding ghost modes \cite{Banerjee:2010qc, Banerjee:2011jp}, around the black hole background. In the path integral approach, the one-loop contribution arises from integrating over these quadratic fluctuations and is formally expressed as the functional determinant of the corresponding kinetic operator.
Although the one-loop partition function plays a central role in capturing quantum corrections to black hole physics, its explicit evaluation in black hole backgrounds is often technically challenging. This difficulty arises primarily from the complexity of determining the spectrum of the associated kinetic operator. The heat-kernel method is a powerful framework to compute the one-loop determinant in the Euclidean formalism \cite{Vassilevich:2003xt}. Alternatively, one can also express the one-loop determinant in terms of quasinormal modes,  which is known as the Denef-Hartnoll-Sachdev formula \cite{Denef:2009kn}. These modes characterize the spectrum of quantum fluctuations and capture the inherently dissipative dynamics of black hole spacetimes. The Denef--Hartnoll--Sachdev (DHS) formula has been quite useful in computing one-loop determinants of fields in black-hole backgrounds in de Sitter and anti-de Sitter spaces \cite{Datta:2011za,Castro:2017mfj,Grewal:2022hlo}, as well as in extracting quantum corrections for extremal and near-extremal black holes \cite{Mukherjee:2024nhx,Kapec:2024zdj,Arnaudo:2024bbd}. In several subsequent works \cite{Datta:2011za,Castro:2017mfj,Anninos:2020hfj,Law:2022zdq,Grewal:2022hlo,Mukherjee:2024nhx}, it has been emphasized that the Denef–Hartnoll–Sachdev prescription works very well in computing the one-loop determinant of scalar fields. For spinning fields, however, the one-loop path integral naturally factorizes as \cite{Anninos:2020hfj}
\begin{align}
    \mathcal{Z}^{(1)}=\mathcal{Z}_{\rm ch}\times \mathcal{Z}_{\rm edge},
\end{align}
where \(\mathcal{Z}_{\rm edge}\) denotes the edge partition function associated with a localized path integral on the codimension-two surface, naturally identified with the bifurcation surface in Lorentzian geometry.

In this work, we focus on evaluating the one-loop determinant of gravitons in the Nariai geometry. The Nariai geometry arises as a special limit of the Schwarzschild--de Sitter black hole, where the black hole and the cosmological horizons coincide \cite{1950SRToh..34..160N}. In this limit, the geometry of the near horizon region takes the form of a direct product, \( dS_2 \times S^2 \). Interestingly, this structure closely resembles the near-horizon geometry of extremal black holes in asymptotically flat or $AdS$ spacetimes, where a decoupled throat region of the form \( AdS_2 \times S^2 \) emerges. However, unlike the extremal limit, the Nariai limit does not rely on the presence of charge or angular momentum. Instead, it stems purely from the existence of a positive cosmological constant and the degeneracy of the black hole and cosmological horizons.

Using the Denef-Hartnoll-Sachdev prescription, we evaluate the bulk partition function of gravitons 
in Nariai geometry in $D=4$ dimensions, which admits the following integral form:
\begin{align}
    \log\mathcal{Z}^{\rm grav}_{\rm ch}
    &=
    \int_0^{\infty}\frac{dt}{2t}
    \frac{1+e^{-t}}{1-e^{-t}}
    \sum_{n=0,l=2}^{\infty}
    2(2l+1)
    \left(
    e^{-t(n+\frac{1}{2}+i\nu_{2, l})}
    +
    e^{-t(n+\frac{1}{2}-i\nu_{2, l})}
    \right)
    \nonumber\\
    &=
    \int_0^{\infty}\frac{dt}{2t}
    \frac{1+e^{-t}}{1-e^{-t}}
    \sum_{l=2}^{\infty}
    2(2l+1)
    \frac{
    e^{-\frac{t}{2}-i\nu_{2, l}t}
    +
    e^{-\frac{t}{2}+i\nu_{2, l} t}
    }{1-e^{-t}},
    \label{tensorch}
\end{align}
where $\nu_{2, l}=\sqrt{(l+2)(l-1)-\frac{1}{4}}$.
The degeneracy factor $2(2l+1)$ accounts for the two independent parity branches or, equivalently, the two physical  polarizations of the graviton, together with the azimuthal degeneracy $-l\leq m\leq l$.
The quasinormal mode construction naturally computes the character partition function associated with the physical propagating graviton modes.

However, the Euclidean one-loop determinant of the graviton on $S^2\times S^2$ captures not only the character partition function associated with the physical graviton modes, but it also includes the additional contribution from the edge modes, which is localized on the codimension-2 surface \cite{Anninos:2020hfj, Grewal:2022hlo, David:2022jfd}. The contribution of the edge modes in the partition function can be written in the following way (the details are given in section \ref{edgepart}.):
\begin{equation}\label{logzedge}
 \log \mathcal{Z}_{\rm edge}=\log \frac{\Omega_2}{\Omega_1}+\int_0^{\infty}\frac{ds}{2s} \, e^{-\frac{\epsilon^2}{4s}}  K_{\rm edge}(s),
\end{equation}
where
\begin{equation}
K_{\rm edge}(s)=-2K_{1,\rm edge}(s)-6K_{0,\rm edge}(s).
\end{equation}
Here \(K_{1,\rm edge}(s)\) is the heat kernel associated with a transverse spin-1 field on \(S^2\), while \(K_{0,\rm edge}(s)\) is the heat kernel of a massless scalar field on \(S^2\), with the zero mode removed. The factor \(\Omega_2\) denotes the contribution from the negative tensor mode on \(S^2\times S^2\), whereas \(\Omega_1\) is the isometry factor, which in the present case is associated with \(\mathrm{vol}(SO(3))\times \mathrm{vol}(SO(3))\). Therefore, the edge partition function naturally admits an interpretation as a localized path integral on the codimension-two surface, involving lower-spin edge degrees of freedom: a transverse spin-1 field and massless scalar fields on \(S^2\).

 The organization of the paper is as follows. In section \ref{sec2}, we briefly review the Denef-Hartnoll-Sachdev prescription for a minimally coupled scalar field. Using the DHS prescription, we write the character partition function from quasinormal modes. In section \ref{sec3}, we present the quasinormal modes of gravitational perturbations, and with those, we derive the quasinormal bulk character in \ref{sec4}. Finally, in section \ref{sc6}, we show the bulk-edge split of the one-loop determinant on $S^2\times S^2$ and offer some remarks on the edge characters.
\paragraph{Note:} While this paper was in preparation, the preprint \cite{Law:2025yec} appeared on arXiv, showing the bulk-edge split of one-loop partition function of the graviton in the Nariai geometry in arbitrary dimension.
\section{Denef-Hartnoll-Sachdev (DHS) \label{sec2}prescription}\label{sec2}

In this section, we briefly review the Denef-Hartnoll-Sachdev (DHS) prescription \cite{Denef:2009kn} to evaluate the one-loop determinant of a scalar field from an infinite product of quasinormal modes.  In the DHS framework, one can think of the one-loop determinant as a meromorphic function of the mass parameter $m^2$ .{ \footnote{In AdS/dS black-hole backgrounds, the mass parameter can often be traded for the conformal dimension $\Delta$ of the corresponding boundary operator; for a generic black-hole background, we  keep $m^2$ as the analytic parameter.} To understand this more explicitly, let us consider a complex scalar field  in the black hole background. The one-loop determinant can be written as
\begin{align}
    \mathcal{Z}^{(1)}&=\int D\phi\, e^{-\int \sqrt{g}\phi^{*}\left(-\nabla^2+m^2\right)\phi}\sim \frac{1}{\det\left(-\nabla^2+m^2\right) }
\end{align}
We analytically continue $\mathcal{Z}^{(1)}(m^2)$ in a complex $m^2$-plane. The one-loop partition function $\mathcal{Z}^{(1)}(m^2)$ can be completely characterized in terms of its poles, zeros, and its value at asymptotic infinity. In particular, we encounter a pole when the determinant is zero. This occurs when $m^2$ is tuned in such a way that there exists a zero mode of the scalar field $\phi$. Let us denote the zero modes by $\phi_{*,k}$. The Euclidean zero modes $\phi_{*,k}$ correspond to solutions of the Klein-Gordon equation that are regular and periodic in Euclidean time. The index $k$ labels the Matsubara (thermal) modes associated with the compact Euclidean time circle, while $*$ represents all additional quantum numbers characterizing the spatial and internal structure of the mode.

The DHS prescription relies on  the connection between the Euclidean zero modes and the Lorentzian quasinormal modes (related via Wick rotation). As it is well known, the quasinormal modes are defined as the solutions of the equation of motion that satisfy the ingoing boundary condition at the horizon. We denote their frequencies by $\omega_{*}(m^2)$, which depend parametrically on the mass squared $m^2$. When $m^2$ is adjusted so that $\omega_{*}(m^2)$ coincides with a Matsubara frequency $\omega_k = 2\pi i k T$, the corresponding Lorentzian mode $\phi_{*,\omega}$ analytically continues to the Euclidean zero mode $\phi_{k,*}$. This matching of mode spectra underlies the spectral representation of the one-loop determinant in terms of quasinormal frequencies.
From the location of poles at $\omega_*(m^2)=\omega_k$, the one-loop can be written as \cite{Denef:2009kn,Datta:2011za,Castro:2017mfj}:
\begin{align}
   \mathcal{Z}_{\rm ch}&=\prod_{k,*} \left(\omega_k-\omega_*(m^2)\right)^{-1}\label{DHS},
\end{align}
where $\mathcal{Z}_{\rm{ch}}$ denotes the one-loop partition function computed from the infinite product over quasinormal modes, which we refer to as the quasinormal character partition function. \footnote{Generically, the DHS expression also contains a holomorphic factor \(e^{\mathrm{Pol}(m^2)}\), where \(\mathrm{Pol}(m^2)\) is a polynomial in \(m^2\). This factor is associated with the UV-divergent part of the determinant, including the logarithmic divergence when the spacetime dimension is even. This function is determined by comparing the \(m^2\to \infty\) asymptotics of the DHS product with the corresponding heat-kernel expansion. Since this factor does not affect the quasinormal-mode character \(\mathcal{Z}_{\rm ch}\), we will formally drop it in the present discussion.}.
\subsection{Character partition function from quasinormal modes}
We study the infinite product over quasinormal modes to evaluate the free energy. While this approach captures many important aspects—such as the ultraviolet structure of the one-loop determinant—it may not fully account for certain features of the theory, including the gauge group volume and the dependence on coupling constants. Keeping these limitations in mind, we now turn to the character partition function of a real minimally coupled free scalar, which is constructed from the infinite product over (anti-)quasinormal modes.
\begin{align}
\mathcal{Z}_{\rm{ch}} 
&= \prod_{\omega_*, \bar{\omega}_*} \prod_{k=-\infty}^{\infty} 
\left(| k| + \frac{i \omega_*}{2\pi T} \right)^{-\frac{1}{4}} \left( |k| - \frac{i \bar{\omega}_*}{2\pi T} \right)^{-\frac{1}{4}} ,
\end{align}
Following the discussion in \cite{Anninos:2020hfj,Law:2022zdq}, we note that when the theory is PT-symmetric, the anti-quasinormal frequencies are not independent of the quasinormal frequencies. In particular, $\bar{\omega}_*$ can be related to $\omega_*$ , i.e, $\bar{\omega}_*=-\omega_*$. This follows from the Lorentzian equation of motion, which is invariant under time reversal \(t\to -t\). Thus, for every quasinormal mode with frequency \(\omega_*\), there is a corresponding anti-quasinormal mode with frequency \(-\omega_*\). Consequently, one may either write the product only over the quasinormal modes with exponent $-\frac{1}{2}$, or keep both the quasinormal and anti-quasinormal products, in which case each factor should be assigned exponent $-\frac{1}{4}$.

While taking the product over the quasinormal modes, we also have to take into account the degeneracy of other quantum numbers present in the quasinormal modes.
To evaluate the free energy from the product over quasinormal modes, we take the logarithm of the product and use the following integral representation of the logarithm \footnote{In principle, one should keep a regulator $\epsilon$ in the lower limit of the integral to keep track of the UV divergence}:
\begin{align}
    -\log x=\int_0^{\infty}\frac{dt}{t}\, e^{-tx}.
\end{align}
We evaluate the character free energy, or the logarithm of the character partition function, in the following way:
\begin{align}
\log Z_{\text{ch}} 
&= \int_0^\infty \frac{dt}{4t} \sum_{\omega_*}\sum_{\bar{\omega_*}} \sum_{k=-\infty}^{\infty}\left( 
 \, e^{-\left( |k| + \frac{i \omega_*}{2\pi T} \right) t}+ \, e^{-\left( |k| - \frac{i \bar{\omega_*}}{2\pi T} \right) t}\right) \\
&= \int_0^\infty \frac{dt}{4t} \frac{1 + e^{-t}}{1 - e^{-t}} \sum_{\omega_*}\sum_{\bar{\omega_*}}\left(
 \, e^{-\left(   \frac{i \omega_*}{2\pi T} \right) t}+ \, e^{\left( + \frac{i \bar{\omega_*}}{2\pi T} \right) t}\right)\nonumber\\
&=\int_0^\infty \frac{dt}{2t} \frac{1 + e^{-t}}{1 - e^{-t}}\chi_{\rm{QNM}}.
\label{qnmdef}
\end{align}
Here \(\chi_{\rm QNM}=\sum_{\omega_*}e^{-i\frac{\omega_*}{2\pi T}t}\) denotes the quasinormal character. In the final line, we have used the PT relation \(\bar\omega_*=-\omega_*\). In the second line, we have performed the sum over Matsubara modes. Thus, the character partition function admits a simple integral representation in terms of quasinormal modes \cite{Grewal:2022hlo,Mukherjee:2024nhx}. This particular form of the integral is very useful for obtaining the bulk-edge decomposition of the one-loop partition function of spinning fields. As we have already mentioned after the equation \eqref{DHS}, this character partition function does not capture information about the coupling constant or the volume of the gauge group;
one must include them separately to write the complete path integral. In the following sections, we compare the quasinormal character partition function with the complete one-loop partition function on $S^2\times S^2$, and identify the bulk and edge contributions of the full determinant.
\section{Nariai geometry}\label{sec3}
The Nariai geometry emerges as a special limit of the Schwarzschild--de Sitter (SdS) spacetime when the black-hole and cosmological horizons approach each other \cite{1950SRToh..34..160N}.

The Schwarzschild–de Sitter metric in four dimensions is given by
\begin{equation}
    ds^2 = -f(r)\, dt^2 + \frac{dr^2}{f(r)} + r^2 d\Omega_2^2,
\end{equation}
where
\begin{equation}
    f(r) = 1 - \frac{2M}{r} - \frac{\Lambda}{3} r^2,
\end{equation}
and \( d\Omega_2^2 = d\theta^2 + \sin^2\theta\, d\phi^2 \) is the metric on the round unit 2-sphere.

For \(0<9M^2\Lambda<1\), the equation \(f(r)=0\) has two positive roots. We denote the smaller root by \(r_b\), corresponding to the black-hole horizon, and the larger root by \(r_c\), corresponding to the cosmological horizon, i.e. \(0<r_b<r_c\). Explicitly, they are given by \footnote{The trigonometric form of the roots, which can easily be obtained by substituting $r=\frac{2}{\sqrt{\Lambda}}\cos\theta$, after which the cubic equation reduces to $\cos3\theta+3M\sqrt{\Lambda}=0$.}
\begin{equation}
    \begin{split}
        r_b&=\frac{2}{\sqrt{\Lambda}}\cos\left[\frac{1}{3}\cos^{-1}\left(-3 M \sqrt{\Lambda}\right)-\frac{2\pi}{3}\right]\\
        r_c&=\frac{2}{\sqrt{\Lambda}}\cos\left[\frac{1}{3}\cos^{-1}\left(-3 M \sqrt{\Lambda}\right)\right]
    \end{split}
\end{equation}

The static Lorentzian region is \(r_b<r<r_c\). Outside this region, \(f(r)<0\), so \(t\) becomes spacelike and \(r\) becomes timelike.

When the black-hole and cosmological horizons coincide, \(r_b\to r_c\), we approach the Nariai limit. In this limit, \(f(r)\) develops a double zero. We denote the degenerate horizon by \(r=r_N\) and, the condition \(f'(r_N)=0\) gives
\begin{equation}\label{doubleroot}
    M=\frac{\Lambda\, r_N^3}{3}.
\end{equation}
Plugging \eqref{doubleroot} into \(f(r_N)=0\), we find
\begin{equation}\label{rN}
    r_N=\frac{1}{\sqrt{\Lambda}},
    \qquad
    M=\frac{1}{3\sqrt{\Lambda}}.
\end{equation}
Combining these relations, we find that the Nariai limit corresponds to
\begin{equation}
    9M^2\Lambda=1.
\end{equation}
The near-horizon geometry obtained in this limit is \(dS_2\times S^2\), with
\begin{equation}
    \ell_N=\frac{1}{\sqrt{\Lambda}},
    \qquad
    R=\frac{1}{\sqrt{\Lambda}}.
\end{equation}
Zooming in to the near-horizon region, this geometry takes a direct product form : $dS_2\times S^2$ \cite{1950SRToh..34..160N}. 
In this region it is useful to introduce new  coordinates $(\tau,\rho)$ for the $dS_2$ factor.  In these coordinates the Nariai metric takes the direct-product form
\begin{equation}\label{nmet}
ds^2
=-f_N(\rho)d\tau^2+\frac{d\rho^2}{f_N(\rho)}+R^2 d\Omega_2^2,
\qquad
f_N(\rho)=1-\frac{\rho^2}{\ell_N^2}.
\end{equation}

Here $\ell_N$ is the de Sitter length of $dS^2$ and $R$ is the radius of the sphere $S^2$ and the Hawking temperature $T_N=\frac{1}{2\pi\ell_N}$.
The Nariai solution is notable for being an exact solution to the four-dimensional Einstein equations with a positive cosmological constant. 

Moreover, the Nariai geometry bears a striking resemblance to the near-horizon geometry of extremal black holes in asymptotically flat or anti-de Sitter (AdS) spacetimes, which typically exhibit a \( \text{AdS}_2 \times S^2 \) throat. In contrast, the Nariai limit does not require electric/magnetic charge or angular momentum. The existence of this near-horizon geometry follows solely from the presence of a positive cosmological constant and the degeneracy of black hole and cosmological horizons.
\subsection{Quasinormal modes in Nariai geometry}
Quasinormal modes in Nariai geometry can be studied analytically. Since geometry is a direct product of $dS_2\times S^2$, wave equations take a simple form in this background.  The spectrum of scalar, vector, and tensor type perturbations on the angular part ($S^2$) are exactly known \cite{Camporesi:1994ga}, which simplifies the computation by a large amount. Essentially, the wave equation becomes a radial equation on $dS_2$ with the additional mass coming from the modes on $S^2$.
The master equation in the tortoise coordinate takes a simple form:
\begin{align}
     \left[\partial_{r_*}^2+\omega^2-V_s(r_*)\right]\psi_s(r)=0,
\end{align}
where $r_*$ is the tortoise coordinate and is defined by the following equation
\begin{align}
    r_*=\int^{r}\frac{d\tilde{r}}{f(\tilde{r})}=\ell_N\,\tanh^{-1}\left(\frac{r}{\ell_N}\right).
\end{align}
This equation maps $r=\pm \ell_N$ to $r_* \to \pm \infty$.

The form of the potential is different for different type of perturbations. We present the solution of the wave equation with the appropriate boundary conditions in the appendix \ref{app1}. 
The quasinormal modes for the massless spin-s ($s\leq 2$) perturbation in the Nariai background can also be computed analytically \cite{Cardoso:2003sw,Venancio:2020ttw}. Here we just present the quasinormal modes, the details of the derivations are presented in \ref{app1}. For a minimally coupled free massless scalar on the Nariai geometry, the quasinormal modes are given by \cite{Cardoso:2003sw,Venancio:2020ttw}
\begin{align}
     \,  \frac{\omega_*}{2\pi T_N}&=-i\left(n+\frac{1}{2}\right)+\nu_{0, l},\label{QNMs}
\end{align}
where $\nu_{0, l}$ is given by $
\nu_{0, l}=\sqrt{l(l+1)-\frac{1}{4}}.$  

In \cite{Cardoso:2003sw,Venancio:2020ttw}, it was shown that gravitational perturbations in the Nariai geometry decompose into two independent parity branches, namely even and odd modes, corresponding to the two polarizations of gravitons with identical quasinormal spectrum. which is given by \cite{Cardoso:2003sw,Venancio:2020ttw} 
\begin{align}
  \,  \frac{\omega_*}{2\pi T_N}&=-i\left(n+\frac{1}{2}\right)+\nu_{2, l},\label{QNMmodes}
\end{align}
where $\nu_{2, l}$ is given by
\begin{align}
\nu_{2, l}=\sqrt{(l+2)(l-1)-\frac{1}{4}}.\label{nus}
\end{align}
 Therefore, both branches must be included in computing the determinant from the quasinormal modes.
\section{Quasinormal bulk-partition function}\label{sec4}
The DHS prescription is quite useful in computing the one-loop partition function of scalar fields. However, in subsequent works \cite{Datta:2011za, Castro:2017mfj, Anninos:2020hfj, Law:2022zdq, Grewal:2022hlo}, it has been shown that the naive DHS formula does not capture the complete one-loop partition function for spinning quantum fields. In general, the one-loop partition function takes the following product form:
\begin{align}
    \mathcal{Z}^{(1)}&=\mathcal{Z}_{\rm ch}\times \mathcal{Z}_{\rm edge},
\end{align}
where $\mathcal{Z}_{\rm ch}$ corresponds to the character partition function obtained from the quasinormal modes, and $\mathcal{Z}_{\rm edge}$ corresponds to the edge partition function, often interpreted as the localized path integral over the codimension-2 surface.

In this section, we study the one-loop character partition function of gravitons in the Nariai geometry.  For notational simplicity, we work in units \(\ell_N=R=1\). Equivalently, all mode frequency eigenvalues are measured in units of the Nariai scale. We will restore $\ell_N$ in the UV divergent section, through the redefinition of the integral variable.

From the quasinormal modes given in \eqref{QNMmodes}, we can derive the character partition function associated physical graviton modes, accounting for both even and odd parity branches \cite{Cardoso:2003sw,Venancio:2020ttw}. Plugging these quasinormal modes into the formula \eqref{qnmdef}, we obtain
\begin{align}
    \log\mathcal{Z}^{\rm grav}_{\rm ch}
    &=
    \int_0^{\infty}\frac{dt}{2t}
    \frac{1+e^{-t}}{1-e^{-t}}
    \sum_{n=0,l=2}^{\infty}
    2(2l+1)
    \left(
    e^{-t(n+\frac{1}{2}+i\nu_{2, l})}
    +
    e^{-t(n+\frac{1}{2}-i\nu_{2, l})}
    \right)
    \nonumber\\
    &=
    \int_0^{\infty}\frac{dt}{2t}
    \frac{1+e^{-t}}{1-e^{-t}}
    \sum_{l=2}^{\infty}
    2(2l+1)
    \frac{
    e^{-\frac{t}{2}-i\nu_{2, l}t}
    +
    e^{-\frac{t}{2}+i\nu_{2, l} t}
    }{1-e^{-t}}.
    \label{tensorch}
\end{align}

The degeneracy factor $2(2l+1)$ accounts for the two independent parity branches or, equivalently, the two physical  polarizations of the graviton, together with the azimuthal degeneracy $-l\leq m\leq l$.
The quasinormal mode construction naturally computes the character partition function associated with the physical propagating graviton modes.

For propagating modes of the graviton in the bulk geometry, we consider the transverse traceless degrees of freedom. In the following section \ref{sc6}, we compute the corresponding character partition function from the Euclidean perspective using the eigenspectrum of the Laplacian.
\section{One-loop determinant of a scalar  on $S^2\times S^2$}
As a warm-up, we begin by evaluating the one-loop determinant of a scalar  field on $S^2\times S^2$ using the heat-kernel approach. We will follow closely \cite{Anninos:2020hfj} to obtain an integral representation of the kernel. Finally, we will match the one-loop determinant of a massless scalar with the quasinormal bulk partition function, presented in \ref{app2}.

The one-loop partition function of a minimally coupled free massless scalar field on $S^2\times S^2$ is given by
\begin{align}
    \log\mathcal{Z}&=-\frac{1}{2}\log\det{}'\Delta_0.
\end{align}
Here, $\Delta_0=-\nabla^2$ is the scalar Laplacian on $S^2\times S^2$ and the prime is used to denote the operator without the zero mode. This mode requires a separate treatment, which we do not carry out here. Since the scalar computation is only used as a warm-up comparison with the quasinormal character, we keep the determinant primed and leave the zero-mode contribution outside the present analysis.

The eigenspectrum of the scalar Laplacian on $S^2\times S^2$ is given by \cite{Volkov:2000ih}
\begin{align}
    \lambda^{(0)}_{n,l}&=n(n+1)+l(l+1),\quad n,l\geq 0\nonumber\\
    d^{(0)}_{n,l}&=(2n+1)(2l+1).
\end{align}
The scalar Laplacian has a genuine zero mode at $(n,l)=(0,0)$. In computing the determinant, we will perform the sum over all modes and then subtract the contribution from the zero-mode.

The heat kernel corresponding to the scalar eigenspectrum can be evaluated as
\begin{align}
 -\frac{1}{2}\log\det{}'\Delta_0
 &=
 \int_0^{\infty}\frac{ds}{2s}
 e^{-\frac{\epsilon^2}{4s}}
 \sum_{\substack{n,l\geq0\\(n,l)\neq(0,0)}}
 d^{(0)}_{n,l}\, e^{-\lambda^{(0)}_{n,l}s}.
 \label{scalrkernel}
\end{align}

In general, the sum over $n$ or $l$ is difficult to perform because the eigenvalues are quadratic in $n$ and $l$. Following \cite{Anninos:2020hfj}, we use the Hubbard--Stratonovich trick to linearize the sum
\begin{equation}
\sum_{n=0}^{\infty} (2n+1) e^{-s (n + \tfrac{1}{2})^2} 
= \int_{C} \frac{du}{\sqrt{4\pi s}} e^{-\frac{u^2}{4s}} f(u).
\end{equation}

The contour $C$ runs from $-\infty$ to $\infty$, slightly above the real axis by a small positive imaginary constant. The function $f(u)$ is given by
\begin{align}
    f(u)=\sum_{n=0}^{\infty}(2n+1)e^{iu(n+\frac{1}{2})}=\frac{e^{\frac{i u}{2}} \left(1+e^{i u}\right)}{\left(-1+e^{i u}\right)^2}.
    \label{fdef}
\end{align}
We plug $f(u)$ into \eqref{scalrkernel} and perform the integral over $s$
\begin{align}\label{scz}
   -\frac{1}{2}\log\det{}'\Delta_0
   =
   \sum_{l=0}^{\infty}(2l+1)
   \int_{C} \frac{du}{2\sqrt{u^2 + \epsilon^2}}
   \left( e^{-\nu_{0, l} \sqrt{u^2 + \epsilon^2}} f(u) \right)
   -\int_0^{\infty}\frac{ds}{2s}\,e^{-\frac{\epsilon^2}{4s}} ,
\end{align}
where
\begin{align}
    \nu_{0, l}=\sqrt{l(l+1)-\frac14}.
\end{align}
The second term of \eqref{scz}, comes from the subtraction of the zero-mode contribution in the determinant.
Note that the above integral in \eqref{scz} has a branch cut from $u=i\epsilon$ along the imaginary axis. We deform the contour $C$ to $C'$, which runs on both sides of the branch cut on the imaginary axis, and substitute $u=it$. We take $\epsilon\rightarrow 0$ at the end to obtain
\begin{align}
    -\frac{1}{2}\log\det{}'\Delta_0
    &=
    \sum_{l=0}^{\infty}(2l+1)
    \int_{0}^{\infty}
    \frac{dt}{2t}
    \frac{1+e^{-t}}{1-e^{-t}}
    \frac{e^{-\frac{t}{2}+i\nu_{0, l}t}+e^{-\frac{t}{2}-i\nu_{0, l} t}}{1-e^{-t}}
    -\int_0^{\infty}\frac{ds}{2s}\,e^{-\frac{\epsilon^2}{4s}} .
    \label{scdet}
\end{align}
The first term in \eqref{scdet} agrees with the scalar quasinormal character in (B.2). The last term in \eqref{scdet} is the explicit subtraction of the constant scalar zero mode, which is not captured by the quasinormal-mode character.
 \footnote{Shown in appendix-\ref{app2}}.
\section{One-loop determinant of graviton on $S^2\times S^2$}\label{sc6}
The gauge fixed one-loop path integral of graviton on $S^2\times S^2$ is given by \cite{Volkov:2000ih}
\begin{align}
    \mathcal{Z}^{1-\rm{loop}}_{\rm{grav}}&=\frac{\Omega_2}{\Omega_1}\frac{\det(\Delta_{(1)})^{\frac{1}{2}}}{\det(\Delta_{(2)})^{\frac{1}{2}}},\label{grone}
\end{align}
where $\Delta_2$ is operator for the transverse traceless tensor fluctuation:
\begin{align}
    \Delta_2 h_{\mu\nu}=-\nabla^2h_{\mu\nu}-2R_{\mu\nu\alpha\beta}h^{\alpha\beta}.
\end{align}
The vector operator acts on the co-exact vector $\xi_{\mu}$:
\begin{align}
    \Delta_1:=-\nabla_{\mu}\nabla^{\mu}-1.
\end{align}
Here, $\Omega_2$ is the contribution from a negative tensor mode on $S^2\times S^2$. $\Omega_1$ is the isometry factor which in this case would be $\rm{vol}(SO(3))\times \rm{vol}(SO(3))$. The details of the gauge fixing procedure along with the counting zero modes and determining path integral measure has been shown in \cite{Volkov:2000ih}. The complete bulk-edge partition function includes the character partition function along with the isometry factor $\Omega_1$. In the following section, we will focus on evaluating the heat-kernels of the determinants isolating the bulk and the edge parts of the character partition function.

\subsection{Eigen spectrum of Laplacians}
From \eqref{grone}, the  spectrum of the operator $(\Delta_{(2)})$ acting on the transverse traceless graviton on $S^2\times S^2$ are given by \cite{Volkov:2000ih}
 \paragraph{Tensor spectrum:}
  \begin{align}
    \lambda^{(2a)}_{n,l} &= n(n+1) + l(l +1)-2, \quad n\geq 2, l\geq 2\nonumber\\
    d_{n,l}&=5(2n+1)(2l+1)
    \label{tens1}
  \end{align}
  In addition to this, we have
\begin{align}
     \lambda^{(2b)}_{n}&=n(n+1),\quad n\geq 2\nonumber\\
      d_n&=18(2n+1)\label{tens2}
\end{align}
And finally, we have another set of eigenvalues' spectrum.
\begin{align}
    \lambda^{(2c)}_{n}&=n(n+1)-2,\quad n\geq 2\nonumber\\
      d_n&=2(2n+1)\label{tens3}
\end{align}
In addition to the three different set of eigen spectrum of the tensor Laplacian, we have nine tensor eigen modes with eigenvalue $2$ and a single tensor mode with a negative eigenvalue $-2$.
  \paragraph{Vector spectrum}
Let us also write down the eigen spectrum of the operator $(\Delta_{(1)})$ and the operator \cite{Volkov:2000ih}.
  \begin{align}
    \lambda^{(1a)}_{n,l} &= (n(n+1) - 1) + (l(l + 1) - 1),\quad n\geq 1, l\geq 1\nonumber\\
     d_{n,l}&=3(2n+1)(2l+1),
   \label{vecs1}.
  \end{align}
In addition to this spectrum, we also have
\begin{align}
     \lambda^{(1b)}_{n} &= (n(n+1) - 2) ,\quad n\geq 2, \nonumber\\
     d_n&=2(2n+1)\label{vecs2}
\end{align}

It is important to note that the tensor spectrum $\lambda^{(2,c)}_{n}$ is the same as the vector spectrum $\lambda^{(1b)}_{n}$, that is, $\lambda^{(2,c)}_{n}=\lambda^{(1b)}_{n}$. Due to this isospectral condition, we get a cancelation between these terms in the heat kernel.

\subsection{One-loop determinant on $S^2\times S^2$}\label{edge}
Let us examine the one-loop determinant from the tensor eigen spectrum in \eqref{tens1}. The heat-kernel corresponding to the tensor eigen modes is given by
\begin{align}
 &\int_0^{\infty}\frac{ds}{2s}e^{-\frac{\epsilon^2}{4s}}\,K_{(2a)} (s) \nonumber\\&=\int_0^{\infty}\frac{ds}{2s}e^{-\frac{\epsilon^2}{4s}}\sum_{n=2,l=2}^{\infty}d_{n,l}\, e^{-\lambda^{(2a)}_{{n,l}}s}\nonumber\\
  &=\int_0^{\infty}\frac{ds}{2s}e^{-\frac{\epsilon^2}{4s}}\sum_{l=2}^{\infty}5(2l+1)\sum_{n=2}^{\infty}(2n+1) \, e^{-s\left((n+\frac{1}{2})^2+\nu_{2, l}^2\right)}\nonumber\\
  &=\int_0^{\infty}\frac{ds}{2s}e^{-\frac{\epsilon^2}{4s}}\sum_{l=2}^{\infty}5(2l+1)\Big(\sum_{n=0}^{\infty}(2n+1) \, e^{-s\left((n+\frac{1}{2})^2+\nu_{2, l}^2\right)}-\sum_{n=0}^{1}(2n+1) \, e^{-s\left((n+\frac{1}{2})^2+\nu_{2, l}^2\right)}\Big)\nonumber\\
  &=\int_0^{\infty}\frac{ds}{2s}e^{-\frac{\epsilon^2}{4s}}\left[\sum_{l=2}^{\infty}5(2l+1)\Big(\hat{K}_{(2a)}(s;l)-K^{(2a)}_{~\rm low}(s;l)\Big)\right],
  \label{tensordet}
\end{align}
where $\nu_{2, l}$ is defined in \eqref{nus}. $\hat{K}_{(2a)}(s;l)$ and $K_{\rm low}(s;l)$ are defined by
\begin{equation}
\begin{split}
\hat{K}_{(2a)}(s;l)&=\sum_{n=0}^{\infty}(2n+1) \, e^{-s\left((n+\frac{1}{2})^2+\nu_{2, l}^2\right)}\\
K^{(2a)}_{~\rm low}(s;l)
&=
\sum_{n=0}^{1}(2n+1)\,
e^{-s\left((n+\frac{1}{2})^2+\nu_{2,l}^2\right)} .
\end{split}
\end{equation}
In the third line, we added the modes with \(n=0\) and \(n=1\) and subsequently subtracted them from the original kernel. We denote this low-mode contribution by \(K_{\rm low}(s;l)\). Note that \(K_{\rm low}(s;l)\) still carries \(l\)-dependence, and therefore must be summed over \(l\) together with the degeneracy factor \(5(2l+1)\). Now we would like to pay special attention to the first term (with all modes from $n=0$ to $\infty$). 

In general, it is difficult to perform the sum over $n$, since it is quadratic in the exponential. However, using the Hubbard-Stratonovich trick, one can linearize the exponential and perform the sum. The linearized sum over $n$ is evaluated in \eqref{fdef}. Following the same trick as shown for the scalar,
we plug the $f(u)$ into \eqref{tensordet} and perform the integral over $s$ to obtain
\begin{align}
   \int_0^{\infty}\frac{ds}{2s}e^{-\frac{\epsilon^2}{4s}}\,K_{(2a)} (s) &=
    \sum_{l=2}^{\infty}5(2l+1)\Big(\int_{C} \frac{du}{2\sqrt{u^2 + \epsilon^2}} \left( e^{-\nu_{2,l} \sqrt{u^2 + \epsilon^2}} f(u) \right)\nonumber\\
    &- \int_0^{\infty}\frac{dt}{2t}\sum_{n=0}^{1}(2n+1) e^{-t\left((n+\frac{1}{2})^2+\nu_{2, l}^2\right)}\Big)
\end{align}
In the second line, we performed the integral over $s$. Let us also deform the contour from $C$ to $C'$ and collect the contribution through the branch cut (with the branch cut point $u=i\epsilon$) and substitute $u=it$. Finally, we take the $\epsilon\rightarrow 0$ limit to obtain
\begin{align}
  \int_0^{\infty}\frac{ds}{2s}e^{-\frac{\epsilon^2}{4s}}\,K_{(2a)} (s) &=\sum_{l=2}^{\infty}5(2l+1)\Big(\int_{0}^{\infty} \frac{dt}{2t}\frac{1+e^{-t}}{1-e^{-t}}\frac{e^{-\frac{t}{2}+i\nu_{2, l}t}+e^{-\frac{t}{2}-i\nu_{2, l} t}}{1-e^{-t}}\nonumber\\&-\int_0^{\infty}\frac{dt}{2t}\sum_{n=0}^{1}(2n+1) e^{-t\left((n+\frac{1}{2})^2+\nu_{2, l}^2\right)}\Big)\label{finaldettens}
\end{align}

We can write \eqref{finaldettens}, in a compact form by performing a single sum over $n=2$ to $\infty$, which is given by
\begin{align}
   \int_0^{\infty}\frac{ds}{2s}e^{-\frac{\epsilon^2}{4s}}\,K_{(2a)} (s) &=\int_0^{\infty}\frac{dt}{2t}\sum_{l=2}^{\infty}5(2l+1)\left(e^{-i\nu_{2, l} t}+e^{i\nu_{2, l} t}\right) \frac{ e^{-\frac{3 t}{2}} \left(5 e^t-3\right)}{\left(e^t-1\right)^2} \label{1ht}
\end{align}
The complete integral representation of the heat-kernel from the eigen spectrum of the tensor modes in \eqref{1ht} agrees with equation (C.7) of \cite{Anninos:2025ltd}. However, we wish to separate the terms with modes $n=0$ and $n=1$, which will enter into the edge partition function.

Similarly, we can evaluate the heat-kernel formula for the transverse vector spectrum in \eqref{vecs1}. Performing similar steps as given for the tensor modes, we find
\begin{align}
  \int_0^{\infty}\frac{ds}{2s}e^{-\frac{\epsilon^2}{4s}}\,K_{(1a)} (s) &=\sum_{l=1}^{\infty}3(2l+1)\Big(\int_{0}^{\infty} \frac{dt}{2t}\frac{1+e^{-t}}{1-e^{-t}}\frac{e^{-\frac{t}{2}+i\nu_{1, l}t}+e^{-\frac{t}{2}-i\nu_{1, l} t}}{1-e^{-t}}\nonumber\\&-\int_0^{\infty}\frac{dt}{2t} e^{-t\left((\frac{1}{2})^2+\nu_{1, l}^2\right)}\Big).\label{finaldetvecs}
\end{align}
Here $\nu_{1, l}=\sqrt{l(l+1)-\frac{9}{4}}$
and again, we added the $n=0$ mode in the kernel and subtracted it off. 

We also note that, 
\begin{align}\label{nurel}
    \nu_{2,l}=\sqrt{(l+2)(l-1)-\frac{1}{4}}=\sqrt{l(l+1)-\frac{9}{4}}=\nu_{1,l}
\end{align}
Since the tensor and vector determinants come with opposite signs in the free energy (or in $\log Z$), we can combine the first term of \eqref{finaldettens} and the first terms of \eqref{finaldetvecs} restricted to $l\geq 2$, and find
\begin{align}
&\int_0^{\infty}\frac{ds}{2s}e^{-\frac{\epsilon^2}{4s}}\,\left[\sum_{l=2}^{\infty}5(2l+1)\,\hat{K}_{(2a)}(s;l)-K_{(1a)}\left(l\geq 2\right)\right]\nonumber\\
    &=
    \int_0^{\infty}\frac{dt}{2t}
    \frac{1+e^{-t}}{1-e^{-t}}
    \sum_{n=0,l=2}^{\infty}
   \left( 5(2l+1)-3(2l+1)\right)
    \left(
    e^{-t(n+\frac{1}{2}+i\nu_{2, l})}
    +
    e^{-t(n+\frac{1}{2}-i\nu_{2, l})}
    \right)
    \nonumber\\
    &=
    \int_0^{\infty}\frac{dt}{2t}
    \frac{1+e^{-t}}{1-e^{-t}}
    \sum_{l=2}^{\infty}
    2(2l+1)
    \frac{
    e^{-\frac{t}{2}-i\nu_{2, l}t}
    +
    e^{-\frac{t}{2}+i\nu_{2, l} t}
    }{1-e^{-t}}.
    \label{tensorchsec6}
\end{align}
In the first line, $K_{(1a)}\left(l\geq 2\right)$ denotes the vector kernel but restricted to $l\geq 2$ modes.
The physical graviton character obtained independently in \eqref{tensorch} is reproduced by the overlapping $l\geq 2$ sector of the Euclidean tensor and vector sectors given in \eqref{finaldetvecs}.

As we have already mentioned, the isospectral condition of the tensor and vector modes $\lambda^{(2c)}_{n}=\lambda^{(1b)}_{n}$ simplifies the rest of the computation of the heat kernel. It is clear from the eigen modes and the degeneracy, in \eqref{tens2} and \eqref{vecs2}, we get a cancelation of the heat kernels of the tensor and vector eigen modes mentioned above in the heat kernel, i.e, $K_{(2c)}-K_{(1b)}=0$.

We are now left with the eigenspectra in \eqref{tens2}. The heat-kernel corresponding to this eigen spectrum is given by
\begin{align}
 \int_0^{\infty}\frac{ds}{2s}e^{-\frac{\epsilon^2}{4s}}\, K_{(2b)}(s)&=\int_0^{\infty}\frac{ds}{2s}e^{-\frac{\epsilon^2}{4s}}\sum_{n=2}^{\infty}18(2n+1)\, e^{-\lambda^{(2b)}_{{n}}s}
\end{align}
Using the Hubbard-Stratonovich trick, we can perform the sum over $n$-modes. 
\begin{align}
\int_0^{\infty}\frac{ds}{2s}e^{-\frac{\epsilon^2}{4s}}\, K_{(2b)}(s)&=\int_0^{\infty}\frac{dt}{2t} \frac{18 e^{-\frac{3 t}{2}} \left(5 e^t-3\right)}{\left(e^t-1\right)^2}(e^{-i\frac{t}{2}}+e^{i\frac{t}{2}})
\end{align}
In addition , to these kernels we have nine constant eigen modes of $2$ and one negative eigen mode $-2$. 

\subsection{Edge partition function}\label{edgepart}

Let us now recombine the complete heat-kernel in the following way:

\begin{align}
K_{1-\rm loop}(s)&= K_{(2a)}(s)-K_{(1a)}(s)+K_{(2b)}(s)+K_{\rm {finite}}(s),
\end{align}
where the finite kernel $K_{\rm {finite}}(s)$ consists of nine constant eigen modes of $2$ and one negative eigen mode $-2$ and isolated tensor modes discussed around \eqref{tensordet}.

As we have already noted in \eqref{tensorchsec6}, the first term of \eqref{tensordet} and the first term of \eqref{finaldetvecs}, but only restricted to the overlapping $l\geq 2$ sector, gives the quasinormal bulk graviton character as presented in \eqref{tensorch}. Precisely, we can write it as
\begin{align}
    K_{\rm ch}^{\rm grav}(s)&=\sum_{l=2}^{\infty}2(2l+1)\sum_{n=0}^{\infty}\left[(2n+1) \, e^{-s\left((n+\frac{1}{2})^2+\nu_{2, l}^2\right)}\right]
\end{align}

We  define the rest as the edge-kernel in the following way:

    \begin{align}\label{kedgeor}
K_{\rm edge}(s)&=K_{1-\rm loop}(s)-K_{\rm ch}^{\rm grav}(s)\nonumber\\
&=-K_{\rm low}^{(2a)}(s)+K_{\rm low}^{(1a)}(s)\big|_{l\geq2}-K_{(1a)}(s)\big|_{l=1}+K_{(2b)}(s)+K_{\rm finite}(s).
\end{align}
Here $K_{\rm low}^{(2a)}(s)$ denotes the kernel for the remaining $n=(0,1)$-low modes in the tensor kernel presented in \eqref{tensordet}. $K_{\rm low}^{(1a)}(s)\big|_{l\geq2}$ denotes the second term in \eqref{finaldetvecs}, but restricted to the $l\geq 2$ sector. Note that the relative sign in these two kernels comes from the fact that, in the full determinant, the tensor and the vector ghost modes come with opposite signs.
\begin{align}
K_{\rm low}^{(2a)}(s)
&=
\sum_{l=2}^{\infty}5(2l+1)
\sum_{n=0}^{1}(2n+1)
e^{-s\left((n+\frac12)^2+\nu_{2,l}^2\right)},\\
K_{\rm low}^{(1a)}(s)\big|_{l\geq2}
&=
\sum_{l=2}^{\infty}3(2l+1)
e^{-s\left(\frac14+\nu_{1,l}^2\right)} .
\end{align}

Using the fact, $\nu_{1,l}=\nu_{2,l}$ as given in \eqref{nurel}, the $n=0$ part of $K_{\rm low}^{(2a)}(s)$ combines with $K_{\rm low}^{(1a)}(s)\big|_{l\geq2}$ to give
\begin{align}\label{kedge1}
-K_{\rm low}^{(2a)}(s)|_{n=0}+K_{\rm low}^{(1a)}(s)\big|_{l\geq2}&=-2\sum_{l=2}^{\infty}(2l+1)e^{-s(l(l+1)-2)} \nonumber\\
&=-2 K_{1,\rm edge}(s)
\end{align}
Note that this term of the kernel is nothing but a transverse-vector edge kernel on $S^2$, and therefore, we denote this as $K_{1,\rm edge}(s)$. 

We now have the remaining $n=1$-term of $K_{\rm low}^{(2a)}(s)$, which contributes to
\begin{align}
    K_{\rm low}^{(2a)}(s)|_{n=1}=15\sum_{l=2}^{\infty}(2l+1)e^{-s l(l+1)}
\end{align}
and $l=1$-vector ghost tower, which contributes to
\begin{align}
    -K_{(1a)}(s)|_{l=1}=-9\sum_{n=1}^{\infty}(2n+1)e^{-s n(n+1)}=-9\sum_{l=1}^{\infty}(2l+1)e^{-s l(l+1)}
\end{align}
and $K_{(2b)}(s)$ which is also given by
\begin{align}
    K_{(2b)}(s)&=+18\sum_{l=2}^{\infty}(2l+1)e^{-s l(l+1)}
\end{align}
The remaining scalar-type terms are
\begin{align}\label{kedge2}
&- K_{\rm low}^{(2a)}(s)|_{n=1}- K_{(1a)}(s)|_{l=1}+K_{(2b)}(s)+9\,e^{-2s}\nonumber\\
&=-15\sum_{l=2}^{\infty}(2l+1)e^{-s l(l+1)}
-9\sum_{l=1}^{\infty}(2l+1)e^{-s l(l+1)}
+18\sum_{l=2}^{\infty}(2l+1)e^{-s l(l+1)}+9\,e^{-2s}\nonumber\\
&=-6\sum_{l=1}^{\infty}(2l+1)e^{-s l(l+1)}=-6K_{0,\rm edge}(s)
\end{align}
The $9\,e^{-2s}$-term comes from the nine tensor modes with eigen value $+2$.
Note that this kernel is precisely the massless scalar-kernel on $S^2$, but the $(l=0)$ zero mode is removed.

From \eqref{kedgeor}, \eqref{kedge1} and \eqref{kedge2}, we find the total edge-kernel
\begin{align}\label{kedgefin}
    K_{\rm edge}(s)&=-2 K_{1,\rm edge}(s)-6K_{0,\rm edge}(s)
\end{align}

Note that the complete path integral in \eqref{grone} also includes the isometry factor $\Omega_1$ and the contribution from the negative  mode, which has not yet been included in \eqref{kedgefin}. Including these, we find the complete edge partition function
\begin{equation}\label{logzedge}
  \log \mathcal{Z}_{\rm edge}=\log \frac{\Omega_2}{\Omega_1}+\int_0^{\infty}\frac{ds}{2s} \, e^{-\frac{\epsilon^2}{4s}}  K_{\rm edge}(s)
\end{equation}
where $K_{\rm edge}(s)$ is given in \eqref{kedgefin}.

It is important to note that the edge partition function in \eqref{logzedge} admits a natural interpretation as a localized path integral over lower-spin fields, namely a transverse spin-1 field and a massless scalar field on \(S^2\), with the scalar zero mode removed.
\paragraph{Consistency check with \cite{Law:2025yec}:}

Here, we  discuss the consistency check with \cite{Law:2025yec}. In equation (4.26) of \cite{Law:2025yec}, it is shown that
\begin{equation}
    \begin{split}
        K(\tau)&=K_{\rm bulk}(\tau)+K_{\rm edge}(\tau)\\
        K_{\rm edge}(\tau)&=-2\left(\sum_{l=-1}^{\infty\prime}D^d_{l,1}e^{-\lambda^{(1)}_{0,l}\tau}+D^3_{1,0}\sum_{l=1}^{\infty}D^d_{l,0}e^{-\lambda^{(0)}_{1,l}\tau}\right)
    \end{split}
\end{equation}

For $d=3$, $D^3_{l,1}=2l+1$, $D^3_{1,0}=3$, and $D^3_{l,0}=2l+1$. Here, $\sum_{l=-1}^{\infty\prime}$ denotes the omission of $l=1$ for which $\lambda^1_{~0,l=1}=0$. Therefore, the edge-kernel of \cite{Law:2025yec} can be written as
\begin{align}\label{kedgelaw}
     K_{\rm edge}(\tau)&=-2\left(\sum_{l=-1}^{\infty\prime}D^d_{l,1}e^{-\lambda^{(1)}_{0,l}\tau}+D^3_{1,0}\sum_{l=1}^{\infty}D^d_{l,0}e^{-\lambda^{(0)}_{1,l}\tau}\right)\nonumber\\
     &=-2\left(\sum_{l=2}^{\infty}(2l+1)\,e^{-\lambda^{(1)}_{0,l}\tau}+3\sum_{l=1}^{\infty}(2l+1)\,e^{-\lambda^{(0)}_{1,l}\tau}\right)+2
\end{align}

In the second line, we have explicitly expanded the sum from $l=2$ to $\infty$ for the first term and isolated the contribution of the $l=-1$-term for which $D^3_{l=-1,1}=-1$ and $\lambda^{(1)}_{0,l=-1}=0$.

We note that the terms in the bracket of \eqref{kedgelaw}, without the analytically continued $l=-1$ contribution, agree precisely with our edge kernel in \eqref{kedgefin}. The additional $+2$ term in \eqref{kedgelaw} arises from the $l=-1$ analytic continuation of the vector edge sector in \cite{Law:2025yec}. This term is not an ordinary vector harmonic, but it has been used to package the negative tensor mode sector. In our path-integral organization, we have kept it separated through the negative-mode factor $\Omega_2$, as in \eqref{logzedge}.

Let us now focus on the edge-partition function, which includes the isometry factor $\log \Omega_1=\log\left[\rm{vol}(SO(3))\times \rm{vol}(SO(3))\right]$. Therefore, with this isometry factor, together with \eqref{kedgefin} and the negative tensor sector, we find our edge-partition function to be the same as that given in equation (4.34) of \cite{Law:2025yec} for $d=3$.

Similarly, the bulk partition function in $d=3$, given in equations (4.32) and (4.33) of \cite{Law:2025yec}, agrees with ours \eqref{tensorch} with the substitution of $\nu_{0,l}=\nu_{1,l}=\nu_{2,l}=\sqrt{(l+2)(l-1)-\frac{1}{4}}$ and $\ell_N=1$. Note that, in this case $I=2$ vanishes on $S^2$ for \cite{Law:2025yec}, indicating the presence of the two parity branches.

\subsection{UV divergences}

The one-loop determinants, or the partition functions, are, in general, UV divergent and require regularization to obtain finite results. In this section, we study the UV divergent structure, in particular the logarithmic divergent terms of \eqref{tensorch} and \eqref{logzedge}. Besides the logarithmic divergence, there are power law divergences as well, which typically need a counter-term to remove it. It can be understood in the following way: if the kernel $K(s)$ admits the following expansion in the $s\to 0$ limit,
\begin{align}
    K(s)&=\frac{a_{-2}}{s^2}+\frac{a_{-1}}{s}+a_0+ O(s)+\cdots,
\end{align}
the logarithm of the partition function can be written as
\begin{align}
    \log Z&=\int_{0}^{\infty}\frac{ds}{2s}\, e^{-\frac{\epsilon^2}{4s}} \, K(s)\nonumber\\
    &=c_4\frac{\ell_N^4}{\epsilon^4}+c_2\frac{\ell_N^2}{\epsilon^2}+\alpha\,\log\frac{\ell_N}{\epsilon}+\cdots
\end{align}

Here, in the second line, we redefined $s=\ell_N^2 \sigma$, and by looking at the integral, one can easily derive the scaling form of the partition function. The leading and subleading divergent terms can be removed by using counter-terms, and in general, the finite term depends on the regularization scheme; however, the logarithmic divergent term is independent of the choice of the regularization scheme.

Let us now compute the logarithmic divergent term for the bulk-partition function in \eqref{tensorch}. Following the appendix (C) of \cite{Anninos:2020hfj}, we can write the bulk-partition function in an expansion around $t=0$
\begin{align}
    \log\mathcal{Z}^{\rm grav}_{\rm ch}&= \int_0^{\infty}\frac{dt}{2t}
    \frac{1+e^{-t}}{1-e^{-t}}
    \sum_{l=2}^{\infty}
    2(2l+1)
    \frac{
    e^{-\frac{t}{2}-i\nu_{2, l}t}
    +
    e^{-\frac{t}{2}+i\nu_{2, l} t}
    }{1-e^{-t}},\nonumber\\
    &= \int_0^{\infty}\frac{dt}{2t}\, F_{\rm ch}^{\rm grav}(t)
\end{align}
Here $F_{\rm ch}^{\rm grav}(t)=\frac{e^{-\frac{t}{2}}(1+e^{-t})}{(1-e^{-t})^2}\sum_{l=2}^{\infty}4(2l+1)\cos(\nu_{2,l}t)$. Following, equations (C.5)-(C.13) of \cite{Anninos:2020hfj}\footnote{Or equivalently, one can perform the $l$-sum using Euler–Maclaurin method, near $t=0$.} and expanding the integrand around $t=0$, we find the constant term of $F_{\rm ch}^{\rm grav}(t)|_{\rm const}=\frac{44}{45}$. Therefore, the log divergent piece of the bulk partition function is given by $ \log\mathcal{Z}^{\rm grav}_{\rm ch}|_{\rm log div}=\frac{22}{45}\log\frac{\ell_N}{\epsilon}$.

In the edge partition function, there are potentially two sources of the log divergences: 1. from the character partition function(including the negative tensor mode in $\Omega_2$, 2. the isometry factor $\Omega_1$. Following the same principle of bulk character, the logarithmically divergent piece of the character partition function of the edge mode is given by $\log \mathcal{Z}^{\rm ch}_{\rm edge}|_{\rm log div}=\left(\frac{22}{3}+2\right)\log\frac{\ell_N}{\epsilon}=\frac{28}{3}\log\frac{\ell_N}{\epsilon}.$

The logarithmic divergence of the isometry factor can be understood from the path integral normalization of the zero modes following \cite{Anninos:2020hfj}. Note that, the regulated determinant is evaluated through the heat-kernel representation
\begin{align}
    \log Z&=\int_0^{\infty}\frac{d\tau}{2\tau}\,e^{\frac{-\epsilon^2}{4\tau}}\sum_i\, e^{-\tau\, \lambda_i}
\end{align}

So, for a single eigen-value $\lambda_i$, the contribution to the path integral is given by
\begin{align}
    \log Z_i&= \int_0^{\infty}\frac{d\tau}{2\tau}\,e^{\frac{-\epsilon^2}{4\tau}}\, e^{-\tau\, \lambda_i}=\log M-\frac{1}{2}\log\lambda_i+O(\epsilon),
\end{align}

where $M=2\frac{e^{-\gamma}}{\epsilon}$ and $\gamma$ is the Euler–Mascheroni constant.

The path integral measure for a canonically normalized scalar field is given by\footnote{In particular, the path integral measure would be $D\phi=\prod_i\frac{M}{\sqrt{2\pi}}d\phi_i$.}
\begin{align}
    ds_{\phi}^2=\frac{M^2}{2\pi}\int (\delta\phi)^2
\end{align}

For the gravitational path integral, the relation between the canonical and the path integral measures, and similarly between the canonical and the path integral group volumes, was established in \cite{Anninos:2020hfj}. 
\begin{align}
ds_{\rm PI}^2
&=
\frac{A_{d-1}}{4G_N\, d(d+2)}
\frac{M^4}{2\pi}
\,ds_c^2,
\\
\frac{{\rm vol}(G)_{\rm PI}}{{\rm vol}(G)_c}
&=
\left(
\frac{A_{d-1}}{4G_N\, d(d+2)}
\frac{M^4}{2\pi}
\right)^{\frac12 \dim G}.
\end{align}

In our case, $\rm \dim G=\rm \dim (SO(3))+\rm \dim (SO(3))=6$, therefore, the $\rm vol (G)_{\rm PI}\sim \left(M\,\ell_N\right)^{12}$. So, $\log\Omega_1\sim 12\log \left(\frac{\ell_N}{\epsilon}\right) $. 

From the above analysis, we find the logarithmic divergent term of the edge partition function is 
\begin{align}\log \mathcal{Z}_{\rm edge}|_{\rm log div}&=\log \mathcal{Z}^{\rm ch}_{\rm edge}|_{\rm log div}-\log\Omega_1|_{\rm log div}\nonumber\\
&=\left(\frac{28}{3}-12\right)\log\frac{\ell_N}{\epsilon}=-\frac{8}{3}\log\frac{\ell_N}{\epsilon}.
\end{align}

\section{Discussion}
In this paper, we evaluate the bulk character partition function of scalars, vectors, and gravitons from their quasinormal mode spectra using the Denef-Hartnoll-Sachdev (DHS) prescription. For minimally coupled scalar fields, we demonstrate that the quasinormal mode character partition function agrees precisely with the nonzero-mode part of the one-loop determinant computed via the heat-kernel method on the $S^2\times S^2$ background. In contrast, for gravitons, the full one-loop determinant obtained from the heat kernel approach does not coincide with the quasinormal mode character partition function. We isolate the edge-character partition function from the heat-kernel expression to account for this difference. We find that the edge partition function  admits a natural interpretation as a localized path integral over lower-spin fields, namely a transverse spin-1 field and a massless scalar field on \(S^2\), with the scalar zero mode removed.

 In \(\mathrm{AdS}\), the Denef-Hartnoll-Sachdev (DHS) prescription \cite{Castro:2017mfj} requires modification for fields with spin \(s > 0\), as certain higher-spin modes become singular at the horizon. Specifically, it has been shown that for a spin-\(s\) field, the modes with the Matsubara frequency number \(|k| < s\) are singular and must be excluded from the DHS product formula. This suggests that, the boundary conditions for higher-spin fields should be treated with greater care near the horizon. It would be interesting to understand the origin of the edge character partition function from the perspective of quasinormal modes in de Sitter black holes.

 We aim to interpret the edge character partition function of gravitons in terms of a putative entropy associated with gravitational edge modes \cite{David:2022jfd,David:2021wrw, Mukherjee:2023ihb, Mukherjee:2021rri}. In four dimensions (\(D = 4\)), it has been demonstrated that the entanglement entropy of the gravitational edge modes \cite{David:2022jfd} matches the edge character partition function computed on \(S^4\) \cite{Anninos:2020hfj}. However, for the case at hand, no analogous independent computation is currently available. We hope to pursue this direction in the near future, which may offer deeper insight into the physical interpretation of the edge character partition function.

\section*{Acknowledgement}
The author thanks Shiraz Minwalla and Justin David for discussions and encouragement. The author also thanks the anonymous referee for helpful comments and suggestions that significantly improved the draft. 

\appendix
\section{Quasinormal modes in Nariai geometry}\label{app1}
The metric of the Nariai space is given in \eqref{nmet}.  In this appendix, we denote
the Nariai static-patch coordinates $(\tau,\rho)$ by $(t,r)$; these should always be understood as coordinates of the $dS_2$ factor in the Nariai geometry, not as the original Schwarzschild--de Sitter coordinates appearing in (3.1).

Let us consider a minimally coupled massless scalar in this background. The wave equation obeyed by the minimally coupled scalar is given by
\begin{align}
    \frac{1}{\sqrt{-g}}\partial_{\mu}\left(\sqrt{-g}g^{\mu\nu}\partial_{\nu}\right)\Phi=0
\end{align}
The wave equation takes a simple Schr\"odinger like form in the tortoise coordinate:
\begin{align}
    \left[\frac{1}{f(r)}\left(\partial_{r_*}^2-\partial_{t}^2\right)+\nabla_{\Omega}^2\right]\Phi=0,\label{waveform}
\end{align}
where $\nabla^2_{\Omega}$ is the scalar Laplacian on $S^2$. For simplicity, we will consider $R=1$.
To solve the wave equation, we can take the following mode ansatz:
\begin{align}
    \Phi(t,r,\Omega)=\psi(r)e^{-i\omega t}Y_{l,m}(\Omega),
\end{align}
where $Y_{l,m}(\Omega)$ is the scalar spherical harmonics on $S^2$. The wave equation \eqref{waveform} can be written in a simple equation of the radial coordinate.
\begin{align}
    \left[\partial_{r_*}^2+\omega^2-V_0(r_*)\right]\psi(r)=0\label{mastereq}
\end{align}
The scalar potential $V_0(r_*)$ is given by
\begin{align}
    V_0(r_*)=\frac{l(l+1)}{\cosh^2(\frac{r_*}{\ell_N})}
\end{align}
The above potential admits a well known  Pöschl-Teller potential \cite{Cardoso:2003sw,Venancio:2020ttw}. The standard technique to solve the equation \eqref{mastereq} is the following:

We first define $\xi$
\begin{align}
    \xi=\frac{1}{2}+\frac{1}{2}\tanh(\frac{r_*}{\ell_N}).
\end{align}
The asymptotic boundary $r_*\rightarrow \infty$ corresponds to $\xi\rightarrow 1$. Let us also define the following parameters
\begin{align}
    a&=\frac{1}{2}+i\sqrt{l(l+1)-\frac{1}{4}},\\
    b&=\frac{1}{2}-i\sqrt{l(l+1)-\frac{1}{4}}\\
    c&=i\omega\ell_N+1.\label{abc}
\end{align}
With the parameters defined in \eqref{abc}, the wave equation takes a simple hypergeometric form in $\xi$ variable under the following redefinition of $\psi(r_*)$:
\begin{align}
\psi(r_*) = \xi^{(c - 1)/2} (1 - \xi)^{\frac{1}{2}(a + b - c)} R(\xi) 
\end{align}
The equation in $R(\xi)$ is given by
    \begin{align}
\xi(1 - \xi) \frac{d^2 R(\xi)}{d\xi^2} + [c - \xi(a + b + 1)] \frac{dR(\xi)}{d\xi} - ab\, R (\xi)= 0. 
\end{align}
The solution of the above equation is given by

    \begin{align}
R(\xi) = A\, F(a, b, c; \xi) + B\, \xi^{1 - c} F(1 + a - c, 1 + b - c, 2 - c; \xi). 
\end{align}
where $F(a,b,c,\xi)$ is the usual Hypergeometric function.
We should impose appropriate boundary conditions to obtain the quasinormal modes. Let us investigate the behavior of the mode functions at $r_*\rightarrow \pm \infty$.
We impose the following boundary conditions: at $r_* \rightarrow -\infty$, the modes should behave like $\sim e^{-i\omega (t+r_*)}$ and at $r_*\rightarrow \infty$, the modes should go like $e^{-i\omega(t-r_*)}$. These set of boundary conditions imply the outgoing waves at the aymptotic infinity and the ingoing waves at the horizon \cite{PhysRevD.30.295}
\begin{align}
\left. \psi(r_*) \right|_{r_* \to +\infty} \rightarrow
&\ B\, \frac{\Gamma(c - a - b)\Gamma(2 - c)}{\Gamma(1 - a)\Gamma(1 - b)}\, e^{-i\omega(t - r_*)} \notag \\
&+ B\, \frac{\Gamma(a + b - c)\Gamma(2 - c)}{\Gamma(a - c + 1)\Gamma(b - c + 1)}\, e^{-i\omega(t + r_*)}.
\end{align}
Therefore, we should reduce the coefficient of $e^{-i\omega(t+r_*)}$ which can be done by setting $a-c+1=-n$ or $b-c+1=-n$, where $n$ is a non-negative integer. This boundary condition is analogous to the boundary condition in the horizon. Choosing the former condition with the definitions in \eqref{abc}, we find
\begin{align}
   \frac{ i\omega_*}{2\pi T_N}=n+\frac{1}{2}+i\nu_{0, l},
\end{align}
where $\nu_{0, l}=\sqrt{l(l+1)-\frac{1}{4}}.$ These set of quasinormal mode frequencies has been computed in \cite{Cardoso:2003sw, Venancio:2020ttw}. 

For gravitational perturbation, the form of the Pöschl-Teller potential type changes, which is mainly due to the different eigen spectrum of vector and tensor modes on $S^2$.
For  the form of the potential are given by
\begin{align}
    V_2(r_*)&=\frac{(l+2)(l-1)}{\cosh^2(\frac{r_*}{\ell_N})}\label{pots}
\end{align}
Following the same procedure as scalar, we can compute the quasinormal modes of the gravitational perturbations. The quasinormal mode frequencies for gravitational perturbations take the following form
\begin{align}
   \frac{ i\omega_*}{2\pi T_N}=n+\frac{1}{2}+i\nu_{2, l},\quad n\geq 0
\end{align}
where $\nu_{2, l}=\sqrt{(l+2)(l-1)-\frac{1}{4}}$. Here, the quantum number $l$ runs from $2$ to $\infty$. 
\section{Scalar quasinormal character}\label{app2}
For completeness, let us also present the quasinormal character partition function of a minimally coupled massless scalar in Nariai geometry.
For a minimally coupled free massless scalar on the Nariai geometry, the quasinormal modes are given by \cite{Cardoso:2003sw,Venancio:2020ttw}
\begin{align}
     \,  \frac{\omega_*}{2\pi T_N}&=-i\left(n+\frac{1}{2}\right)+\nu_{0, l},\label{QNMs}
\end{align}
where $\nu_{0, l}$ is given by $
\nu_{0, l}=\sqrt{l(l+1)-\frac{1}{4}}.$ 

Using the quasinormal modes of scalars, we plug this expression into \eqref{qnmdef} to find,
\begin{align}
    \log\mathcal{Z}^{\rm scalar}_{\rm{ch}}&=\int_0^{\infty}\frac{dt}{2t}\frac{1+e^{-t}}{1-e^{-t}}\sum_{l=0}^{\infty}(2l+1)\frac{e^{-\frac{t}{2}+i\nu_{0, l} t}+e^{-\frac{t}{2}-i\nu_{0, l} t}}{1-e^{-t}}\label{scch}
\end{align}
We note that the scalar quasinormal character agrees with the first term in \eqref{scdet}. The full primed determinant contains, in addition, the explicit subtraction of the constant scalar zero mode, which we do not analyze in this paper.
\bibliographystyle{JHEP}
\bibliography{biblio.bib}
\end{document}